\documentclass[prl,twocolumn,superscriptaddress,nofootinbib,amsmath,amssymb,aps,]{revtex4-2}

\usepackage{graphicx,enumerate,xcolor,enumitem}
\usepackage[colorlinks,citecolor=blue]{hyperref}
\newcommand{\beq}{\begin{equation}}
\newcommand{\eeq}{\end{equation}}
\newcommand{\bea}{\begin{eqnarray}}
\newcommand{\eea}{\end{eqnarray}}
\newcommand{\nn}{\nonumber}
\allowdisplaybreaks[4]

\begin{document}

\title{CP Phases in 2HDM and Effective Potential: A Geometrical View}

\author{Qing-Hong Cao}
\email{qinghongcao@pku.edu.cn}
\affiliation{Department of Physics and State Key Laboratory
of Nuclear Physics and Technology, Peking University, Beijing 100871, China}
\affiliation{Collaborative Innovation Center of Quantum Matter, Beijing, China}
\affiliation{Center for High Energy Physics, Peking University, Beijing 100871, China}

\author{Kun Cheng}
\email{chengkun@pku.edu.cn}
\affiliation{Department of Physics and State Key Laboratory
of Nuclear Physics and Technology, Peking University, Beijing 100871, China}

\author{Changlong Xu}
\email{changlongxu@pku.edu.cn}
\affiliation{Department of Physics and State Key Laboratory
of Nuclear Physics and Technology, Peking University, Beijing 100871, China}

\begin{abstract}
Using a geometric description of 2HDM, we classify CP invariants into three independent sectors such as scalar potential, Yukawa interaction and CKM matrix. Thermal effective potential of 2HDM is calculated in a basis invariant way. It is shown that the CP violation in Yukawa interactions can contribute to effective potential at one loop level but the CP phase in the CKM matrix cannot leak to effective potential at all orders. In the 2HDM with a softly broken $Z_2$ symmetry, the leading thermal correction tends to restore the CP symmetry at high temperature.
\end{abstract}

\maketitle

\noindent{\bf Introduction.~}Two-Higgs-Doublet-Model (2HDM) is one of the simplest extensions of Standard Model (SM) that can provide both new sources of CP violation and strong first order phase transition~\cite{Lee:1973iz,Gunion:2005ja,Branco:2005em,Branco:2011iw,Battye:2011jj,Pilaftsis:2011ed}. 
It suffers, unfortunately, from the arbitrariness of the scalar basis choice, e.g., a unitary transformation between the two Higgs doublets does not have any physical consequence. The CP phases in the 2HDM could originate from the scalar potential, Yukawa interactions, or Cabbibo-Kobayashi-Maskawa (CKM) matrix~\cite{Kobayashi:1973fv}. To fully understand the CP property of the 2HDM, one needs to clarify independent CP invariants. As a global symmetry the CP symmetry is best studied in basis invariant methods, such as the bilinear notation~\cite{THDMbilinear,THDMbilinear_CPV,Ivanov:2006yq,Ivanov:2007de,Nishi:2006tg} or the tensor notation~\cite{JforScalarFermion,Davidson:2005cw,Trautner:2018ipq}. In the Letter we adopt the bilinear notation to categorize all the independent CP invariants through a geometric prospective, which makes the separation of the CP invariants from the Yukawa interactions and the CKM matrix intuitively evident. For the first time, effective potential including the contributions from the scalar self interactions, the Yukawa interactions and the gauge interactions are calculated fully in a way of basis invariant form. 

A general fourth-degree scalar potential is 
\begin{align} 
    &V(\Phi_1,\Phi_2)=m_{11}^2\Phi_1^{\dagger}\Phi_1+m_{22}^2 \Phi_2^{\dagger}\Phi_2
    -m_{12}^2 \Phi_1^{\dagger}\Phi_2 \nonumber\\
    &+\tfrac{1}{2} \lambda_1 (\Phi_1^{\dagger }\Phi_1)^2
    +\tfrac{1}{2} \lambda_2 (\Phi_2^{\dagger }\Phi_2)^2 +\lambda_3 (\Phi_2^{\dagger}\Phi_2) (\Phi_1^{\dagger }\Phi_1)\nonumber\\
    &+\lambda_4 (\Phi_1^{\dagger }\Phi_2) (\Phi_2^{\dagger }\Phi_1)
    +\tfrac{1}{2} \lambda_5 (\Phi_1^{\dagger }\Phi_2)^2\nonumber\\
    &+\lambda _6(\Phi_1^{\dagger }\Phi_1) (\Phi_1^{\dagger }\Phi_2) 
    +\lambda _7 (\Phi_2^{\dagger }\Phi_2) (\Phi_1^{\dagger }\Phi_2) 
    +h.c.~,
\label{eq:V2hdm}     
\end{align}
in which $(m_{12}^2,\lambda_{5,6,7})$ are generally complex while all the others parameters are real. The scalar potential can be reparametrized by an $SU(2)_\Phi$ rotation $\Phi_i' = U_{ij}\Phi_j$ ($i,j=1,2$).
Making use of the relation between $SU(2)$ and $SO(3)$ groups, the basis transformation can be viewed explicitly in the so-called $K$-space, in which the $SU(2)_\Phi$ basis transformation $\Phi_i' = U_{ij}\Phi_j$ corresponds to an $SO(3)_K$ rotation.  Define a four vector $K^\mu\equiv \Phi_i^\dagger \sigma_{ij}^\mu \Phi_j=(K_0,\vec{K})^T$, where
\begin{align}
&\vec{K}= ( \Phi_1^\dagger\Phi_2+\Phi_2^\dagger\Phi_1,i(\Phi_2^\dagger\Phi_1-\Phi_1^\dagger\Phi_2) ,\Phi_1^\dagger\Phi_1-\Phi_2^\dagger\Phi_2)^T,\nn\\
&K_0= \Phi_1^\dagger\Phi_1+\Phi_2^\dagger\Phi_2.
\end{align}
Under an $SO(3)_K$ rotation $R_{ab}(U)= \tfrac{1}{2} \text{tr}\left[ U^\dagger \sigma_a U \sigma_b \right]$, $K_0$ behaves as a singlet while $\vec{K}$ transforms as a vector, i.e.,
\begin{equation}\label{eq:SO3}
  (K_0^\prime,\vec{K}^\prime) = (K_0, R(U) \vec{K}).
\end{equation}
The scalar potential in the $K$-space is~\cite{THDMbilinear,Ivanov:2007de}
\begin{align}
        V&=\xi_0 K_0 + \eta_{00} K_0^2+\vec \xi \cdot \vec{K}
        + 2 K_0 \vec \eta \cdot \vec{K}+ \vec{K}^T {E}\vec{K},
    \label{eq:Vbilinear}
\end{align}
where 
\begin{eqnarray}
\xi_{0} &\equiv& \frac{1}{2}(m_{11}^{2}+m_{22}^{2}), \qquad
\eta_{00} =(\lambda_{1}+\lambda_{2}+2 \lambda_{3})/8\nonumber\\
\vec{\xi} &=& \left(- \Re\left(m_{12}^{2}\right), \Im\left(m_{12}^{2}\right), \tfrac{1}{2}(m_{11}^{2}-m_{22}^{2})\right)^{T},\nonumber\\
\vec{\eta} &=&\left(
\Re(\lambda_{6}+\lambda_{7})/4,-\Im(\lambda_{6}+\lambda_{7})/4, (\lambda_{1}-\lambda_{2})/8\right)^{T},\nonumber\\
E&=&\frac{1}{4}\left(\begin{array}{ccc}
\lambda_{4}+\Re\left(\lambda_{5}\right) & -\Im\left(\lambda_{5}\right) & \Re\left(\lambda_{6}-\lambda_{7}\right) \\
-\Im\left(\lambda_{5}\right) & \lambda_{4}-\Re\left(\lambda_{5}\right) & \Im\left(\lambda_{7}-\lambda_{6}\right) \\
\Re\left(\lambda_{6}-\lambda_{7}\right) & \Im\left(\lambda_{7}-\lambda_{6}\right) & \left(\lambda_{1}+\lambda_{2}-2 \lambda_{3}\right)/2
\end{array}\right).\nonumber
\end{eqnarray}
As the scalar potential is invariant under the $SO(3)_K$ rotation, it demands the coefficients transform covariantly in the dual space of $K^\mu$, i.e. $\xi^\prime_0 = \xi_0$,  $\eta^\prime_{00} = \eta_{00}$,  $
\vec{\xi}^\prime = R(U)\vec{\xi}$, $\vec{\eta}\prime = R(U)\vec \eta$ and $E^\prime = R(U) E R(U)^T$.

\begin{figure}
\includegraphics[width=.4\linewidth]{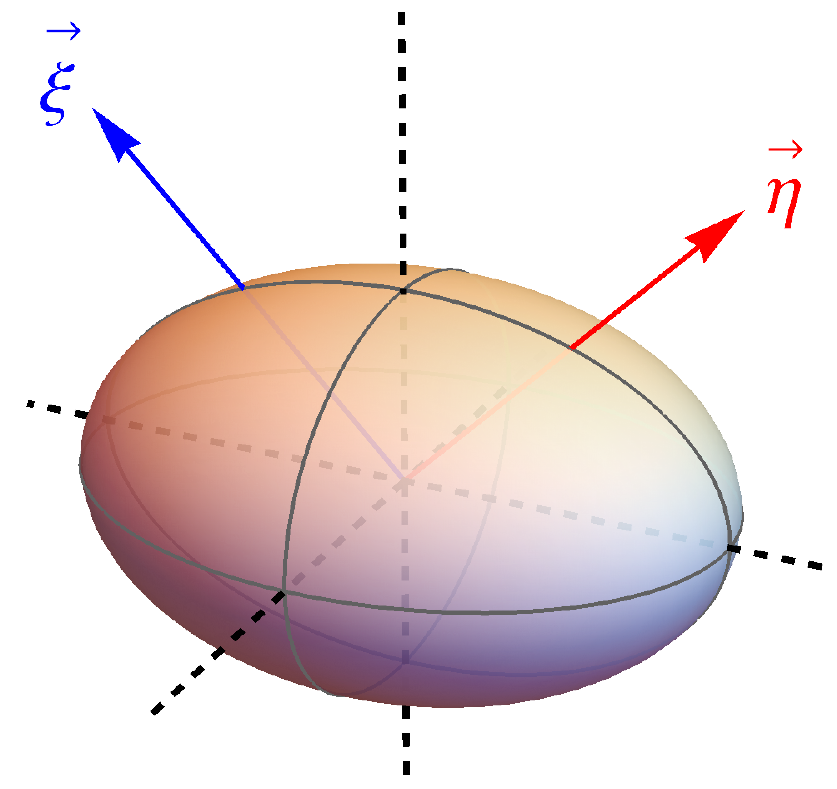}
\caption{Configuration of parameter vectors in a CP symmetric potential, where the black dashed line denotes the three principal axis of the ellipsoid.}
\label{fig:geometricview}
\end{figure}

The conventional CP transformation $\Phi_i \to \Phi_i^*$ corresponds to a mirror reflection $K_2 \to -K_2$ in the $K$-space~\cite{THDMbilinear_CPV}. The CP invariant potential satisfies
\begin{equation}
\label{eq:CPC}
    V(\Phi_1,\Phi_2) = V(\Phi_1,\Phi_2)|_{\Phi_i\to\Phi_i^*},
\end{equation}
while in the $K$-space it becomes
\begin{equation}
    V(K_0,\vec{K})  = V(K_0,\vec{K})|_{K_2 \to -K_2}.
\end{equation}
It requires that $m_{12}$ and $\lambda_{5,6,7}$ are all real.
Denote $\hat{\Pi}$ as the mirror operator that reflects $K_2$ when acting on $\vec{K}$, i.e. $\hat{\Pi} (K_1, K_2, K_3)=(K_1, -K_2, K_3)$. The potential is invariant under the mirror reflection, on condition that
\begin{equation}\label{eq:CPGeometricCondition}
    \vec{\xi}=\hat{\Pi} \vec{\xi},\quad\vec{\eta}=\hat{\Pi} \vec{\eta},\quad E=\hat{\Pi} E \hat{\Pi}^T,
\end{equation}
which can be easily understood from a geometric view. The $3\times 3$ real symmetric tensor $E$ possesses at least three $C_2$ axis (principal axis) and three symmetry plane. In order to respect the CP conserving condition in Eq.~\ref{eq:CPGeometricCondition}, $(E,\vec{\xi},\vec{\eta})$ are all invariant under the mirror reflection, therefore, $\vec{\xi}$ and $\vec{\eta}$ must lie on the same symmetry plane of $E$; see Fig.~\ref{fig:geometricview}. 
The symmetric tensor $E$ can be visualized as a ellipsoid when positive definite or as a hyperboloid when not positive definite. 

\vspace*{3mm}
\noindent{\bf Independent CP invariants.~}Next we explore all the independent CP invariants which are free of the basis choice. The existence of a non-zero CP invariant is a sign of CP violation, and it corresponds to a non-zero CP phase.
The CP symmetry is conserved if and only if all the CP invariants vanish. 

1) Scalar potential: Two independent CP invariants in scalar potential can be constructed from $\vec{\xi}$, $\vec{\eta}$ and $ E$~\cite{THDMbilinear_CPV},
\begin{equation}\label{eq:JforECPV} 
    I_1 = (\vec \xi \times \vec \eta)\cdot E\vec\xi, \quad 
    I_2 = (\vec \xi \times \vec \eta)\cdot E\vec\eta.
\end{equation}
In case of $\vec{\xi}$ is collinear with $\vec{\eta}$ along direction $\vec{l}$, another invariant $(\vec{l}\times E\vec{l})\cdot E^2 \vec{l}$ can be constructed; see Ref.~\cite{THDMbilinear_CPV} for details.

2) Yukawa interaction: For simplicity we consider the quark sector in this work. The Yukawa couplings for quarks in the 2HDM Lagrangian are
\begin{align}\label{eq:LYukawa}
&\mathcal{L}_{\rm Yuk} = -\bar{Q}_L^m y^{mn}_{u,i}\tilde{\Phi}_i u^n_{R} -\bar{Q}_L^m y^{mn}_{d,i}\Phi_i d^n_{R} + h.c.~,
\end{align}
where $Q_L=(u_L,d_L)^T$ and the superscripts $m$ and $n$ sum over all generations.
The Yukawa couplings $y_{u,i}$ and $y_{d,i}$ are in general complex, therefore, the Yukawa couplings $y_{1}$ and $y_{2}$ for a specific quark have four degree of freedom together. One can parameterize them by an overall strength $|\mathcal{Y}|=\sqrt{|y_1|^2+|y_2|^2}$ and three angles as
\begin{equation}
    (y_1,y_2)=\mathcal{Y}(s_{\beta},c_{\beta}e^{-i\gamma})=e^{i\delta} |\mathcal{Y}| (s_{\beta},c_{\beta}e^{-i\gamma}).
\end{equation}
Usually, the Yukawa coupling terms in the Lagrangian cannot be expressed in bilinear notation. One can use the bilinear notation to deal with the Yukawa couplings if the couplings can be shown to be in vector representations of $SU(2)_\Phi$ basis rotation. 
For each individual generation, $y_{u,i}$ and $y_{d,i}^*$ are covariant with $\Phi_i$ under basis transformation $\Phi_i\to U_{ij}\Phi_j$.
We define covariant vectors in the dual space of $K^\mu$ in terms of the Yukawa couplings as follows: 
\begin{equation}\label{eq:YukawaY}
 Y_{u,\mu}\equiv y_{u,i}^* (\sigma_\mu)_{ij} y_{u,j},\quad
 Y_{d,\mu}\equiv y_{d,i} (\sigma_\mu)_{ij} y_{d,j}^*,
\end{equation}
which transform in the same way as $\xi_\mu$ and $\eta_\mu$, i.e. 
\begin{equation}\label{eq:YukawaYvec}
(Y_0,\vec{Y})_{u/d}^{mn}  \to (Y_0, R(U)\vec{Y})_{u/d}^{mn}.
\end{equation}
The definition in Eq.~\ref{eq:YukawaY} yields
\begin{equation}
    \vec{Y}=|\mathcal{Y}|^2(s_{2\beta} c_{\gamma},-s_{2\beta} s_{\gamma}, -c_{2\beta}),\quad Y_0=|\vec{Y}|.
\end{equation}
On the other hand, the common phase of $y_{1}$ and $y_{2}$, $\delta$, is related the CKM matrix and absent in the $SO(3)_K$ vectors.

If the Lagrangian preserves the  CP symmetry, vectors $\vec Y_{u/d}^{mn}$ should be invariant under the mirror reflection $\hat{\Pi}$ and lie on the same reflection plane of $\vec{\xi}$ and $\vec{\eta}$ shown in Fig.~\ref{fig:geometricview}. We construct CP invariants as follows: 
\begin{equation}\label{eq:Jyukawa}
    J_{u/d}^{mn}=(\vec{\xi} \times \vec{\eta})\cdot \vec{Y}_{u/d}^{mn}.
\end{equation}
From the geometrical view, when $\vec{Y}$ is not in the same plane of $\vec{\xi}$ and $\vec{\eta}$, nonzero CP invariant $J$ is generated, and its magnitude is the ``volumn" of the cross product $(\vec{\xi} \times \vec{\eta})\cdot \vec{Y}$. 
Considering the Yukawa interactions of $N$-generation quarks, we obtain $2N^2$ new CP invariants $J_u^{mn}$ and $J_d^{mn}$ where $m,n = 1,2,\cdots, N$.

3) The CKM matrix:
For each complex quark mass matrix $M_u$ and $M_d$, one can always perform a singular decomposition such that $V_u^L M_u V_u^R$ and $V_d^L M_d V_d^R$ are real diagonal matrices, where $V^{L,R}_{u,d}$ stands for a unitary quark basis transformation. But with three families of quarks, the complex phases in $M_u$ and $M_d$ cannot be simultaneously removed with the same $V^L$. Therefore, the CKM matrix $V_{\rm CKM}= (V_u^L)^\dagger V_d^L$~\cite{Kobayashi:1973fv} is left with a phase, which can be described by the so-called Jarlskog invariant~\cite{Jarlskog:1985ht,Dunietz:1985uy}, a basis invariant quantity, 
\begin{equation}
    J=\det[M_u M_u^\dagger, M_d M_d^\dagger].
\end{equation}
The quark mass matrices given by Eq.~\ref{eq:LYukawa} are $M_u^{mn}= y_{u,i}^{mn}v_i^*$ and $M_d^{mn}= y_{d,i}^{mn}v_i$, and each element of quark mass matrices is a $SU(2)_\Phi$ singlet.
Since the quark masses do not exhibit any $SO(3)_K$ tensor structures, the Jarlskog invariant of CKM matrix constructed by $SU(3)_{L/R}$ basis rotations of quark basis is different from those CP invariants constructed by $SU(2)_\Phi$ rotations such as the invariants shown in Eq.~\ref{eq:JforECPV} and Eq.~\ref{eq:Jyukawa}.

To demonstrate the difference between the CP invariants in Yukawa couplings and the Jarlskog invariant in the CKM matrix explicitly, we consider the following real 2HDM potential in the Higgs basis~\cite{Lavoura:1994fv},
\begin{equation}\label{eq:higgsbasis} 
   H_1=\begin{pmatrix}
     G^+\\
     \dfrac{v+\phi+iG^{0}}{\sqrt{2}} 
   \end{pmatrix},~~~~
   H_2=\begin{pmatrix}
     H^+\\
     \dfrac{R+iI}{\sqrt{2}} 
   \end{pmatrix},
\end{equation}
The potential is symmetric under the CP transformation $H_i\to H_i^*$:
\begin{equation} 
    V(H_1,H_2)=V(H_1,H_2)|_{H_i\to H_i^*},
\end{equation}
demanding both $\vec{\xi}$ and $\vec{\eta}$ lie in the plane normal to $\vec{K}_2$,  
therefore $\vec{\xi}\times \vec{\eta}\propto \hat{K}_2$.
The most general Yukawa couplings can be written as 
\begin{align}
    &\sum_{m,n =1}^N \mathcal{Y}_u^{mn} \bar{Q}^m_L\left(  \tilde{H}_1 \sin\beta^{mn}_u + e^{-i\gamma^{mn}_u}\tilde{H}_2 \cos\beta^{mn}_u  \right)u_R^n \nn\\
    +&\sum_{m,n =1}^N \mathcal{Y}_d^{mn} \bar{Q}^m_L\left( H_1 \sin\beta^{mn}_d + e^{i\gamma^{mn}_d}H_2 \cos\beta^{mn}_d \right)d_R^n \nn \\
    +&~~h.c.~,
\label{eq:generalYukawaHiggsbasis}   
\end{align}
which yields $Y_\mu=|\mathcal{Y}|^2(1,s_{2\beta} c_{\gamma} , -s_{2\beta} s_{\gamma}, -c_{2\beta})$ for each generation.
Then CP phase in quark mass matrix is only related to the phase of overall factor $\mathcal{Y}^{mn}_{u/d}$ while CP invariants $J_{u/d}^{mn}$ are only related to $\gamma_{u/d}^{mn}$ and $\beta_{u/d}^{mn}$.

Before going further, we would like to summarize what we have learned so far. Counting the numbers of independent CP invariants in the 2HDM, there are 2 from the scalar potential, $2N^2$ from the Yukawa interactions, and $(N-1)(N-2)/2$ from the CKM matrix. To conserve the CP symmetry, the vectors $\vec\xi$, $\vec\eta$, and $\vec{Y}_{u/d}^{mn}$ have to lie on the same reflection plane of $E$. Each vector acts as an independent CP violation source if it departs from the reflection plane.

\vspace*{3mm}
\noindent{\bf Effective potential.~}Up to now our classification of CP invariants is based only on the tree-level analysis. Next we examine the effective potential which contains quantum corrections from all the interactions mentioned above. Using the background method, the one-loop effective potential at zero temperature is~\cite{Quiros:1999jp}
\begin{equation}
    V_{\rm eff}(\phi_c)=V_{\rm tree}(\phi_c)+V_{\rm CW}(\phi_c),
\end{equation}
where $V_{\rm tree}$ denotes the tree level potential and 
\begin{equation}
    \begin{aligned}
       V_{\rm CW}(\phi_c) &=\frac{1}{2}\mathbf{Tr}\int \frac{d^4p}{2\pi^4}\ln \left[p^2+\mathbf{M}^2(\phi_c)\right]\\
       &= \frac{1}{64\pi^2} \sum_i n_i m_i^4(\phi_c) \left[ \ln \frac{m_i^2(\phi_c)}{\mu^2} - c_i \right].
       \label{eq:Vcw}
 \end{aligned}
\end{equation}
is the Coleman-Weinberg (CW) potential~\cite{Coleman:1973jx} calculated in the Landau gauge under $\overline{\rm MS}$ scheme,  $\mathbf{M}^2$ is the mass matrix with eigenvalues $m_i^2$, $n_i$ denotes the degree of freedom of the field, and $c_i$ is equal to $5/6$ for gauge bosons and $3/2$ for others. 

The effective potential of the 2HDM has been discussed extensively~\cite{Cline:2011mm,Basler:2019iuu,Basler:2016obg,Ferreira:2019bij,Basler:2018cwe}. As a usual practice, only neutral or CP even components of Higgs boson doublets are treated as background fields, which breaks the $SU(2)_L$ invariance explicitly such that our previous discussions cannot apply to $V_{\rm eff}(\phi_c)$.
In order to analyze the CP property of effective potential in bilinear notation, the mass matrix needs to be evaluated in a $SU(2)_L$ invariant way.
For that, we take all the components of the Higgs boson doublets to be background fields, and $\Phi_i=(\phi_{i\uparrow},\phi_{i\downarrow})^T$ should be understood as background fields hereafter.

1) Contributions from the scalar loop: {\it In scenario that the scalar potential preserves the CP symmetry at the tree level, the scalar self interaction cannot induce CP violation effects in the effective potential}. It sounds trivial but hitherto verified only in a specific basis. Below we provide a basis independent proof. 

The mass matrices of the scalar sector cannot be analytically diagonalized, therefore we use the method in Ref.~\cite{Degee:2009vp} to derive the bilinear form of the potential. 
We first expand the trace of logarithm in Eq.~\ref{eq:Vcw} as Taylor series and calculate the trace for each term. For example, it yields 
\begin{align}
\mathbf{Tr}(m^{2}_{S})&=8R_\mu A^\mu + 4S_{\mu\nu}\eta^{\nu\mu}\nn\\
&=\left(20\eta_{00}+4\operatorname{tr}(E)\right)K_0+24\vec{K}\cdot\vec{\eta}+8\xi_0,
\end{align}
where
\begin{align}\label{eq:SSAR}
    S^{\mu\nu}=&R^{\mu}K^{\nu}+R^{\nu}K^{\mu}-g^{\mu\nu}(RK),\nn\\
    S_{A}^{\mu\nu}=&A^{\mu}K^{\nu}+A^{\nu}K^{\mu}-g^{\mu\nu}(AK),\nn\\
    A_{\mu}=&2\eta_{\mu\nu}K^{\nu}+\xi_{\mu},\nn\\
    R^\mu=&(1,0,0,0).
\end{align}
The result is consistent with that in Ref.~\cite{Ivanov:2008er}. Our final result is 
\begin{equation}
V_{CW}^{(S)} = \mathcal{F}\left(S^{\mu\nu} \eta_{\nu\rho} , ~S_A ^{\mu\nu} \eta_{\nu\rho} \right) ,
\end{equation}
where $\mathcal{F}$ is a function of the traces of $S^{\mu\nu} \eta_{\nu\rho}, S_A ^{\mu\nu} \eta_{\nu\rho}$ and their combinations~\cite{CCX_ongoing}.
If the tree level potential is invariant under the mirror reflection $\hat{\Pi}$ operation, i.e. 
\beq
V_{\rm tree}(K_0,\vec{K})= V_{\rm tree}(K_0,\hat{\Pi}\vec{K}),
\eeq
then any combination of the tensors given in Eq.~\ref{eq:SSAR} is invariant too. As a result, the $V_{\rm CW}^{(S)}$ is also CP invariant, i.e.
\beq
V_{\rm CW}^{(S)}(K_0,\vec{K})= V_{\rm CW}^{(S)}(K_0,\hat{\Pi}\vec{K}),
\eeq
as it should be. 

2) Contributions from the gauge boson loop: The mass matrices of gauge bosons can be diagonalized and written in gauge invariant form directly. The eigenvalues of gauge boson masses obtained from the kinetic terms are
\begin{align}\label{eq:WZmass}
   & m^2_Z=\frac{g^2}{8}\left( (1+t_W^2)K_0 + \sqrt{4t^2_W|\vec K|^2+(t_W^2-1)^2K_0^2} \right),\nn\\
    &m^2_\gamma=\frac{g^2}{8}\left( (1+t_W^2)K_0 - \sqrt{4t^2_W|\vec K|^2+(t_W^2-1)^2K_0^2} \right),\nn\\
   & m^2_{W^\pm}=\frac{g^2}{4}K_0,
\end{align}
where $t_W=\tan\theta_W$, and a massless photon is ensured by the neutral vacuum condition $K_0 = |\vec K|$~\cite{THDMbilinear}.
Therefore, the CW potential from gauge boson loop contributions $V_{\rm CW}^{(G)}=V_{\rm CW}^{(G)}(K_0,|\vec K|)$ is spherically symmetric in the $K$-space,
\begin{equation*}
    V_{\rm CW}^{(G)}(K_0,\vec{K})= V_{\rm CW}^{(G)}(K_0,R\vec{K}),\qquad R\in O(3).
\end{equation*}
Furthermore, {\it any global symmetry exhibited by the tree level potential cannot be broken by quantum corrections from gauge bosons}. 

3) Contributions from the quark loop: Usually, the dominant correction of quark loops to the effective potential is from the heaviest quark. Nevertheless, we include both top and bottom quarks to ensure our calculation being $SU(2)_L$ invariant. The top and bottom quark masses can mix as there are charged background fields. The fermion mass matrix derived from $-\partial^2\mathcal{L}/\partial \bar\psi^i_{L}\partial \psi^j_{R} $ reads as
\begin{equation}
    (\bar{t}_L,  \bar{b}_L)
    \begin{pmatrix}
     y_{it}\phi_{i\downarrow}^* & y_{ib}\phi_{i\uparrow} \\
     -y_{it}\phi_{i\uparrow}^* & y_{ib}\phi_{i\downarrow}
    \end{pmatrix}
    \begin{pmatrix}
     t_R\\
     b_R
    \end{pmatrix}.
\end{equation}
After singular decomposition $M_{\rm diag}=L^{-1}MR$, two elements of the diagonalized mass matrix are
\begin{equation}
    m_{t/b}^2= \dfrac{B \pm \sqrt{B^2+C}}{2},
\end{equation}
where $B$ and $C$, in terms of $Y_\mu=(Y_0,\vec{Y})$ defined in Eq.~\ref{eq:YukawaY}, are 
\begin{align}
    B=&\frac{1}{2}(Y_{t0} + Y_{b0} )K_0+ \frac{1}{2}(\vec Y_t + \vec Y_b )\cdot \vec{K}, \nn \\
    C=&-\frac{1}{2}(Y_t\cdot Y_b)K_0^2
    -(Y_{t0}\vec{Y}_b+Y_{b0}\vec{Y}_t)K_0\vec{K}
    \nn \\
    & +\frac{1}{2}\vec{K}\cdot(\vec Y_t\cdot\vec Y_b-Y_{t0}Y_{b0}-\vec Y_t\otimes\vec Y_b-\vec Y_b\otimes\vec Y_t)\cdot\vec{K}.
    \nn
\end{align}
The symbol ``$\otimes$" means direct product of two vectors. For illustration, we consider the special case of $y_t\gg y_b$, in which the top quark plays the leading role. The mass square of top quark is 
\beq
m_t^2=\frac{1}{4}(Y_{t0}K_0+\vec Y_t\cdot\vec K).
\eeq
Consider a scalar potential conserving the CP symmetry at the tree level, i.e. $V_{\rm tree}(K_0,\vec{K})=V_{\rm tree}(K_0,\hat{\Pi}\vec{K})$. When the Yukawa coupling breaks the CP symmetry, or equivalently, $\vec{Y}_{t/b} \neq \hat{\Pi} \vec{Y}_{t/b}$, $m_t^2$ is no longer invariant under the mirror reflection and introduces the CP violation effect to the effective potential at one-loop level,
\begin{equation}
    V_{\rm CW}^{(F)}(K_0,\vec{K})\neq V_{\rm CW}^{(F)}(K_0,\hat{\Pi}\vec{K}).
\end{equation}
The CP violation effect is related to $|J_{t}|$.

4) Contributions from the CKM matrix: Last but not least, we consider the impact of the Jarlskog invariant on the effective potential. Ref.~\cite{Fontes:2021znm} studied the possibility of the Jarlskog invariant entering the effective potential at three loop level, but their calculation of three-loop tadpole diagrams of the CP-odd scalar field shows no impact from the Jarlskog invariant. We notice that the CP violation effect appears in the effective potential only in the form of $SO(3)_K$ tensors. As the Jarlskog invariant cannot be described by the tensor structure of $SO(3)_K$ group, we conjecture that {\it the CP phase in the CKM matrix would not leak into the effective potential through quantum corrections}.

Without loss of generality, consider a CP-conserving Type-I 2HDM in which the $SO(3)_K$ covariant parameter tensors are $E$, $\vec{\xi}$, $\vec{\eta}$ and $\vec{Y}$. To respect the CP symmetry, all the tensors are invariant under a mirror symmetry $\hat\Pi$, i.e.
\begin{align}
    E_{ab}&=E_{cd} \Pi_{ac}\Pi_{bd}, 
    &&\xi_{a}=\xi_{c}\Pi_{ac},\nn\\
    \eta_{a}&=\eta_{c}\Pi_{ac},
    &&Y_{a}=Y_{c}\Pi_{ac}.
\end{align}
Any global $SU(2)_L$ invariant effective potential can be written as
\begin{align}
    &V_{\rm eff}(\Phi_i^\dagger\Phi_j)=V_{\rm eff}(K_0,\vec{K})\nn\\
   =&V_{\rm eff}(K_0,T^{(1)}_a K_a,T^{(2)}_{ab} K_a K_b , T^{(3)}_{abc}K_a K_b K_c , \cdots),
\end{align}
where $T^{(q)}_{a_1\cdots a_q}$ transforms as a rank-$q$ tensor under the $SO(3)_K$ rotation. 
In the $K$-space the tensor $T^{(q)}_{a_1\cdots a_q}$ can be constructed by tensor products of tree level parameter tensors. Moreover, the tensor constructed by $E,\vec{\eta},\vec{\xi},\vec{Y}$ is also invariant under the mirror reflection,
\begin{equation}
    T^{(q)}_{a_1\cdots a_q} = T^{(q)}_{b_1\cdots b_q}\Pi_{a_1 b_1}\cdots \Pi_{a_q b_q}.
\end{equation}
It guarantees the CP invariance of effective potential,
\begin{equation}\label{eq:effCPC}
    V_{\rm eff}(K_0,\vec{K})=V_{\rm eff}(K_0,\hat{\Pi}\vec{K}).
\end{equation}
Even though the quark mass matrix, whose elements are $SU(2)_\Phi$ singlets, may enter $T^{(q)}_{a_1\cdots a_q}$ as a factor, tensor structures of $T^{(q)}_{a_1\cdots a_q}$ remain unchanged and the result of Eq.~\ref{eq:effCPC} still holds.

\vspace*{3mm}
\noindent{\bf Thermal corrections.~}At finite temperature, thermal correction $V_T$ should also be included in the effective potential~\cite{Quiros:1999jp},
\begin{align}
    V_{\rm eff}&=V_{\rm tree}+V_{\rm CW} + V_T,\nn\\
V_{T}&=\sum_i n_i\frac{T^4}{2\pi^2}J_{B/F}\left(m_{i}^{2} / T^{2}\right).
\end{align}
At a high-temperature the leading contributions of the thermal bosonic function $J_{B}$ and fermionic function $J_{F}$ yield 
\begin{align}
    \label{eq:VTG}
    V_{T}^{(G)}&\approx \frac{g^2 T^2}{32}(3+t_W^2)K_0,\\
    \label{eq:VTF}
    V_{T}^{(F)}&\approx -\frac{T^2}{8}\left[(Y_{t0}+Y_{b0})K_0+(\vec{Y}_t+\vec{Y}_b)\cdot\vec K) \right],\\
    \label{eq:VTS}
    V_{T}^{(S)}
    &\approx\frac{T^2}{6}\left[\left(5\eta_{00}+\operatorname{tr}(E)\right)K_0+6\vec{\eta}\cdot\vec{K} \right].
\end{align}
Noted that the leading thermal correction from gauge bosons is $SO(3)_K$ symmetric and will not change any global symmetry in the potential, while contributions from quarks and scalars can modify global symmetries of the potential by shifting tree-level parameters $\vec{\xi}$:
\begin{align}
   \label{eq:vecxiT}
   \vec \xi &\rightarrow \vec \xi + \frac{T^2}{8} \left(8\vec \eta-\vec{Y}_t-\vec{Y}_b\right),
\end{align}
which shows that the CP violations in the quadratic couplings can be affected by the quartic couplings and the Yukawa interactions at high temperature.

\vspace*{3mm}
\noindent{\bf Softly broken $Z_2$ symmetry.~}Finally, we examine the 2HDM with a softly broken $Z$ symmetry which is often studied in literature. The 2HDM often induces flavor changing neutral current which is prohibited by precision measurements. A $Z_2$ symmetry,
\beq
\Phi_1 \to -\Phi_1, \quad  \Phi_2 \to \Phi_2,
\eeq
is therefore introduced to forbid the flavor changing neutral current~\cite{Glashow:1976nt}.
The $Z_2$ symmetry
demands $ m_{12}=\lambda_{6}=\lambda_7=0$. 
The parameter tensors are of the patterns 
\begin{equation}
\vec{\xi} = \left(\begin{array}{c}
0 \\
0 \\
\#
\end{array}\right),~~
\vec{\eta} = \left(\begin{array}{c}
0 \\
0 \\
\#
\end{array}\right),~~
E=\frac{1}{4}\left(\begin{array}{ccc}
\# & \# & 0 \\
\# & \# & 0 \\
0 & 0 & \#
\end{array}\right),
\end{equation}
where the symbol ``$\#$" denotes combinations of other coefficients. From geometrical perspective, $\vec{\xi}$ and $\vec{\eta}$ are on the third primary axis of the symmetric tensor $E$, and the $Z_2$ symmetry is nothing but a $180^\circ$ rotation ($C_2$) around the third primary axis, i.e.  $C_2\vec{K}=C_2(K_1,K_2,K_3)=(-K_1,-K_2,K_3)$. A $Z_2$ invariant 2HDM satisfies  
\begin{equation}\label{eq:Z2GeometricCondition}
    \vec{\xi}=C_2 \vec{\xi},\quad  \vec{\eta}=C_2 \vec{\eta}, \quad E=C_2 E C_2^T. 
\end{equation}
The vectors $\vec{Y}_{u/d}^{mn}$ defined in Eq.~\ref{eq:YukawaYvec} are covariant under an $SO(3)_K$ rotation, therefore also satisfy
\begin{equation}
\vec{Y}_{u/d}^{mn}=C_2\vec{Y}_{u/d}^{mn},
\end{equation}
and it is parallel to $\vec{\xi},\vec{\eta}$ and the $C_2$ axis of $E$ likewise.  $\vec{Y}_{u/d}^{mn}$ points to the same direction in Type-I 2HDM, while $\vec{Y}_u^{mn}$ and $\vec{Y}_d^{mn}$ are in opposite directions in Type-II case. We thus use $\vec{Y}$ to label the direction of $\vec{Y}_{u/d}^{mn}$ for simplicity. Clearly, a $Z_2$ symmetric 2HDM Lagrangian is always CP invariant.

\begin{figure}
    \includegraphics[width=.35\linewidth]{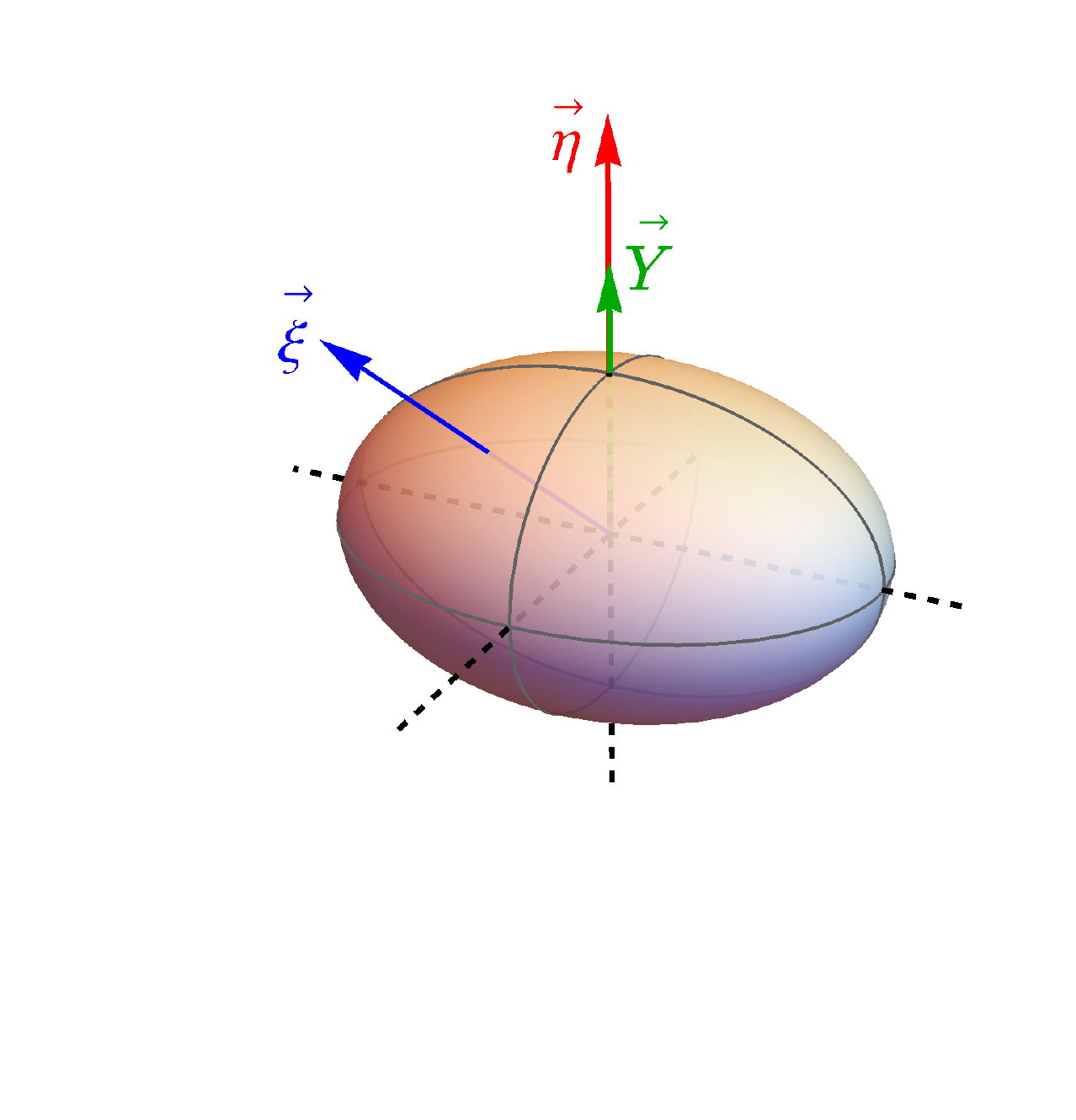}
    \caption{Parameter vectors for a broken $Z_2$ symmetry.}
    \label{fig:softbrokenZ2}
\end{figure}

The $Z_2$ symmetry can be softly broken when $m_{12}^2\neq 0$. The vector $\vec{\eta}$ is not changed, but $\vec{\xi}=(\#, 0, \#)^T$ points to an arbitrary direction; see Fig.~\ref{fig:softbrokenZ2}. Hence, the scalar potential exhibits only one CP invariant, $(\vec \xi \times \vec \eta)\cdot E\vec\xi$, and no CP invariant arises from the Yukawa interactions as $\vec{Y} \parallel \vec{\eta}$. As a result, the 2HDM potential preserving the CP symmetry at tree level still maintain the CP invariance at loop level.
In addition, as shown in Eq.~\ref{eq:vecxiT}, leading thermal correction to $\vec{\xi}$, the only CP violating source, is CP conserving as both $\vec{\eta}$ and $\vec{Y}$ lie on the principal axis of $E$. As long as the length of $\vec{\eta}$ and $\vec{Y}$ are not fine tuned, $\vec \xi$ tends to be bend towards the principal axis at sufficient high temperatures, restoring the $Z_2$ and CP symmetries. 

\vspace*{3mm}
\noindent{\bf Conclusion.~}We generalized the bilinear notation of 2HDM scalar potential to Yukawa couplings by defining dual vectors $\vec{Y}^{mn}_{u/d}$ in bilinear space. By doing so, we obtained all the independent CP invariants in the 2HDM from a geometric view. For the first time, the separation of CP invariants from the Yukawa interactions and the CKM matrix is made intuitively evident. We categorized the CP invariants into three independent sectors, i.e. scalar self interaction, Yukawa interaction and the CKM matrix.

We calculated the Coleman-Weinberg potential in a basis invariant manner. The scalar potential that preserves the CP symmetry at tree level can receive CP violation corrections {\it only} from the Yukawa interactions at one-loop level. We proved that the CP phase in the CKM matrix cannot leak to effective potential at all orders based on the basis invariant form. We further showed that the leading thermal corrections shift the scalar quadratic couplings only with Yukawa and scalar quartic couplings. When a softly broken $Z_2$ symmetry is imposed, the scalar quadratic terms $\vec{\xi}$ is the only source that breaks $Z_2$ and CP symmetries, but its effect tends to be suppressed by Yukawa couplings and scalar quartic couplings at large temperature such that the $Z_2$ and CP symmetries tend to be restored.

\noindent{\bf Acknowledgments.~}We thanks Yandong Liu, Jiang-Hao Yu and Hao Zhang for useful suggestions. The work is supported in part by the National Science Foundation of China under Grant Nos. 11725520, 11675002, and 11635001.

\bibliographystyle{apsrev}
\bibliography{ref}
\end{document}